\documentclass[aps,prl,twocolumn,floatfix]{revtex4}

\usepackage{epsf}
\usepackage{subfigure}
\usepackage{amsmath}
\usepackage{amssymb}
\usepackage{bm}
\usepackage{graphicx}
\usepackage{epstopdf}
\newcommand{\be}{\begin{equation}}
\newcommand{\ee}{\end{equation}}
\newcommand{\bea}{\begin{eqnarray}}
\newcommand{\eea}{\end{eqnarray}}

\newcommand{\parent}[1]{\left( #1 \right)}

\begin{document}

\title{\bf An ergodic process of zero divergence-distance from its time-reversed process}


\author{David Andrieux}

\begin{abstract}
An ergodic process $P$ is constructed such that the divergence-rate $D(P || P^*)$ is zero, yet $P$ is not equal to its time-reversed process $P^*$.
The process $P$ is constructed as a special realization of the universal coding found by Xu \cite{X98}.
This result shows that there exist time-asymmetric processes that are indistinguishable under time-reversal and that can, in principle, be generated with zero dissipation.
\end{abstract}

\maketitle

\vskip 0,25 cm

In this paper, we consider stationary and ergodic processes on a finite set $A$. 
A process $P$ defines a measure on $A^n$, also denoted by $P$, by the formula
\bea
 P(a_1^n) = {\rm Prob}(x : x_1^n = a_1^n) \, ,
\eea
where $a_m^n$ denotes the sequence $a_m,a_{m+1},\ldots,a_n$, with each $a_i\in A$. 

The $n$th-order divergence of a process $P$ with respect to a process $Q$ is defined by
\bea
D_n(P||Q) = \sum_{a_1^n} P(a_1^n) \log\frac{P(a_1^n)}{Q(a_1^n)} \, .
\eea
When it exists, the limit
\bea
D(P||Q) = \lim_{n \rightarrow \infty} \frac{1}{n} \sum_{a_1^n} P(a_1^n) \log\frac{P(a_1^n)}{Q(a_1^n)}
\eea
is called the divergence-rate of $P$ with respect to $Q$.
If $Q$ is a Markov process, $D(P || Q)$ exists for all stationary processes $P$. 
The divergence-rate is always nonnegative. 

The divergence rate plays an important role in coding theory, hypothesis testing theory, and large deviations theory \cite{C84}.
In coding theory, $D_n(P|| Q)$ measures the redundancy or extra cost incurred in using the Shannon code for $Q_n$ on $n$-length sample paths drawn from $P$, rather than the Shannon code for $P_n$. 
In hypothesis testing if the null hypothesis is that a sample path comes from a process $P$, versus the alternative that it comes from $Q$, the divergence-rate is the exponential rate of convergence of the type II error, given a fixed type I error.
In large deviations theory the rate function, which quantifies the probability of observing the statistics $P$ in a process $Q$, is given by $D(P|| Q)$.

Let $P^*$ denote the time-reversed process of $P$, defined by the measure
\bea
P^* \parent{a_1^n} = P \parent{a_n^1} \, .
\eea
We show that there exist processes $P$ such that 
\bea
D\parent{P||P^*}=0 \quad {\rm with} \quad P\neq P^* \, .
\eea 
Note that $P$ cannot be an i.i.d. or a Markov process. 
Indeed, if $Q$ is an i.i.d. or a Markov process and $D(P|| Q) = 0$ then $P=Q$.

\section*{Implications and significance}


If $D(P || P^*) = 0$ then the process $P$ cannot be distinguished from its reversed process $P^*$, at any fixed level, in such a way that the second-kind error goes to $0$ exponentially fast \cite{C84}. 
In other words, we cannot distinguish the arrow of time of this process.

In physics, the breaking of time-reversal is associated to the entropy production under stationary nonequilibrium constraints. 
More precisely, the entropy production of a Markov process $P$ is given by \cite{S76, JVN84, C99, MK03, JQQ04, G04, KPV07, RP12}
\bea
\frac{1}{k_BT} \frac{{\rm d}_iS}{{\rm d}t} =   \frac{1}{2} D_2(P || P^*)=  D(P || P^*) \, ,
\eea
where $k_B$ is the Boltzmann constant and $T$ is the temperature.
This result was verified experimentally using the data of the position of a Brownian particle in a moving optical trap \cite{AGCGJP}.

For non-Markovian processes, plausibility arguments suggest that $D(P||P^*)$ provides a lower bound on the entropy production required to generate the process $P$.
The existence of processes such that $D(P||P^*)=0$ therefore indicates that there exist stationary, time-asymmetric processes that can be generated with arbitrary low dissipation (for a fixed time-asymmetry).

\section*{Construction of the process}

Our construction builds on a remarkable result by Shaogang Xu \cite{X98}. 
Xu constructed an ergodic process $Q$ such that $D(P|| Q) = 0$ {\it for every stationary process $P$}.
In addition, the process $Q$ can be taken to be a B-process, i.e. the stationary coding of an i.i.d. process.

This was the first time a stationary and ergodic process whose Shannon code sequence is universal was shown to exist. 
The consequences of such an universal code are still to being explored. 
As an example, using this universal code it is possible to erase an unknown sequence of data with the minimal achievable dissipation \cite{AG08}.  

We show that the construction of Xu can be adapted to construct a process $Q$ that is time-asymmetric. 
Then, because $D(P|| Q) = 0$ for every stationary process $P$, it will automatically satisfy $D(Q^*||Q) = 0$.
We will first sketch Xu's construction before indicating how to modify it to make the resulting process time-asymmetric.

The process $Q$ is constructed using the method of "cutting and stacking", a method that has been used to construct many examples and counter-examples in ergodic theory.
The cutting and stacking method provides a way to describe sample paths of a stationary ergodic process as concatenations of non-overlapping blocks of varying lengths (Figure 1). 


In essence, the cutting and stacking method cuts the unit interval into subintervals that are assembled using a geometric process. 
This process defines an application $T$ of the unit interval into itself that, combined with a partition $\Pi$ of the unit interval, generates a measure on sequences $A^\infty$. 
Because the Lebesgue measure is preserved during the cutting and stacking process, the resulting application will generate a stationary process, while certain conditions on the asssembly process will garantee ergodicity. 
We refer to references \cite{S91, S98} for an introduction to the method.

The starting point to obtain the process $Q$ of zero divergence is the following observation \cite{S93}.
\\

{\bf Lemma 1}: {\it If two processes $Q$ and $P$ satisfy the condition
\bea
Q (a_1^n) \geq \alpha_n P(a_1^n)\, , \quad a_1^n \in A^n
\label{bound}
\eea
where $\alpha_n$ decreases to $0$ and $(1/n) \log \alpha_n$ goes to $0$, then $D(P||Q) = 0$.}\\

This lemma is at the basis of the following theorem \cite{S98}.\\

{\bf Theorem 1}: {\it Given a countable set $\{ P_i\}$ of regenerative processes, there is an ergodic process $Q$ such that $D(P_i ||Q) = 0$ for all $i$.}\\

Note that if a process is ergodic Markov of some order, then it is regenerative.

The basic idea to demonstrate Theorem 1 is to make $Q$ look like the processes $P_i$ on part of the measure space, so that the lower bound (\ref{bound}) will hold for a given $n$, and make $Q$ look like something different on the rest of the space. 
Then a fraction of the first part, on which $Q$ is mimicking the set of $\{P_i\}$, is mixed into the second part. 
If this fraction is small enough then (\ref{bound}) will hold for $n + 1$; passing to the limit will yield the desired $Q$. 

\begin{figure}[t!]
\centerline{\includegraphics[width=8cm]{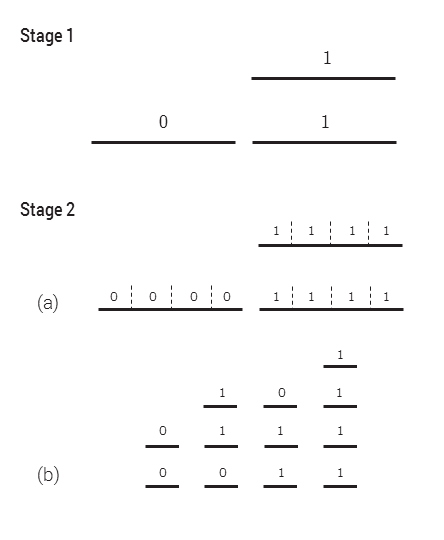}}
\caption{{\bf Example of the cutting and stacking method}. 
In this example, the process is composed of random $0$'s and $1$'s in which $1$'s occur in pairs. 
In the first stage we cut the unit interval into three subintervals of length $1/3$ and form two columns from these.
To go to the second stage, we (a) cut each first stage column into four subcolumns of equal width and (b) stack these subcolumns in pairs in all possible ways.
We go from stage $n$ to stage $n+1$ by cutting each column into $2.2^{2^{n-1}}$ subcolumns of equal width and stacking these subcolumns in pairs into $2^{2^n}$ columns such that all possible combinations of pairs occur \cite{S93}.
}
\label{fig1}
\end{figure}

More precisely, the process $Q$ is generated as follows. Choose a sequence $1 > \beta_n > \alpha_n > 0$ such that
\begin{itemize}  
\item $\alpha_n$ and $\beta_n$ decrease to $0$ as $n\rightarrow \infty$,
\item $(1/n)\log \alpha_n$ and $n\beta_n$ go to $0$ as $n\rightarrow \infty$,
\item $\sum_{n=1}^\infty \beta_n =1$.
\end{itemize}

At each stage $n$ of the process, $Q$ is given by the union of the column structures $S_i$ of the processes $P_i$ with $i\leq n$, plus a 'mixing' component $R(n)$. 
Each column structure $S_i$ is of measure $\beta_n$. 
To go to the next stage of the construction, $R(n+1)$ is generated by repeated cutting and stacking with itself and with copies of size $\beta_{n}-\beta_{n+1}$ of the processes $P_i,$ with $i\leq n$.
At the same time, copies of size $\beta_{n+1}$ of the column structures $S_i$ are repeatedly cut and stacked. 
Finally, the column structure $S_{n+1}$ of size $\beta_{n+1}$ is added to the mix. 
The number of cutting and stacking steps is chosen to garantee ergodicity through mixing and to ensure that the bound (\ref{bound}) holds for all processes $i$. 
Passing to the limit yields the desired $Q$.\\

The final ingredient in the construction is obtained from the fact that any stationary process can be approached by Markov processes in the sense of divergence-rate. 
Let $P^{(j)}$ is the $j$th-order Markovization of $P$ defined by the transition probabilities
\bea
P(x_{j+1} | x_1^j) = \frac{P(x_1^{j+1})}{P(x_1^j)}\, .
\eea
We then have the following result \cite{X98}.
\\

{\bf Lemma 2}: {\it For any stationary process $P$, $\lim_{j\rightarrow \infty} D\parent{P || P^{(j)}} = 0$.}\\

We now take $\{P_i\}$ to be the countable set of all stationary processes that are Markov of some order with rational positive transition probabilities. 
By combining Theorem 1 and Lemma 2, the resulting process $Q$ will have a zero divergence-rate from all stationary processes.\\

From this construction we see that the process $Q$ is not unique, as different (admissible) sequences $\beta_n$ will lead to different processes. 
We will use this freedom to ensure that the resulting process $Q$ is time-asymmetric. 
Since it will also be of zero divergence from all stationary processes, it will follow that $D(Q^*||Q) = 0$.


Assume that $Q$ is symmetric under time reversal. 
Select one symmetric $P_m$ and one asymmetric process $P_r$ in the set $\{P_i\}$ and change the weights $\beta_m$ and $\beta_r$ to $\beta'_m = \beta_m-\delta$ and $\beta'_r = \beta_r + \delta$, where $0< \delta \leq \min(\beta_{r+1}, \beta_{m-1})$.
This ensures that $\sum_n \beta_n =1$ and that $\beta_n$ remains a decreasing sequence. 
The new process $Q'$ will then be composed of more asymmetric (non-overlapping) blocks than $Q$, and therefore will be asymmetric.

Another way to generate a time-asymmetric process is to order the set $\{P_i\}$ in such a way that the time-asymmetric Markov chains come earlier in the set. 
This will assign them greater weights in the construction process and thus generate a time-asymmetric measure.
Indeed, since a process $P_i$ contributes by $\beta_i$ to the total measure and $\beta_n$ is a decreasing sequence, processes that appear early in the set $\{P_i\}$  have larger weigths than processes that appear later in the set.


\vskip 0.5 cm

{\bf Disclaimer.} This paper is not intended for journal publication.


\end{document}